\documentstyle[12pt]{article}

\begin{document}

\author{V.\ Calian\thanks{%
e-mail address: violetaxx@hotmail.com} \\
University of Craiova, Department of Physics,\\
13 A.\ I.\ Cuza, Craiova 1100, Romania}
\title{Quantum BRST operators in extended BRST anti BRST formalism }
\maketitle

\begin{abstract}
The quantum BRST anti BRST operators are explicitly derived and the
consequences related to correlation functions are investigated. The
connection with the standard formalism and the loopwise expansions for
quantum operators and anomalies in Sp(2) approach are analyzed.

\end{abstract}


\section{Introduction}

The most powerful method for covariantly quantizing gauge theories is based
on the antibracket-antifield formalism \cite{1}-\cite{3}. This technique has
been extended by Batalin, Lavrov, Tyutin, in order to include the anti-BRST
symmetry \cite{4}-\cite{7}.

The geometrical interpretation given by Witten in terms of multivectors on
the supermanifold of the fields was also generalized to the Sp(2)-symmetric
formalism, using a description in an overcomplete basis \cite{8} and
introducing further variables ($\overline{\phi }^A$) that kill the
redundancy in cohomology. This allowed one to prove the equivalence of the
two approaches and to express the quantum master equation in extended BRST
anti BRST formalism as the condition that $\exp \left( \frac i\hbar W\right) 
$ is ''divergencefree'', just as in the standard case.

Moreover, the fact that the path-integral does not depend on the choice of
the gauge-fixing function was written in a more elegant way in terms of the
operator $U_K$ and integration theory \cite{8}: 
\begin{equation}
\label{1}\int \left[ D\phi \right] A=\int \left[ D\phi \right] \left(
U_KA\right) _0
\end{equation}
\begin{equation}
\label{2}U_{K^{\prime }}A=U_KA+\overline{\Delta }C
\end{equation}
for some $C$ and if $\overline{\Delta }A=0$, while: 
\begin{equation}
\label{3}U_KA=A+\overline{\Delta }JA
\end{equation}

The contact with the well-known formulation in \cite{4}-\cite{5} is made by
the choice: 
\begin{equation}
\label{3a}U_K=\exp \left[ K,\overline{\Delta }\right] 
\end{equation}
and: 
\begin{equation}
\label{3b}\frac i\hbar K=\frac{\delta F}{\delta \phi ^A}\frac \delta {\delta
\phi _{B1}^{*}}-\frac{\delta F}{\delta \phi ^A}\frac \delta {\delta \phi
_{B2}^{*}} 
\end{equation}
for a gauge-fixing boson $F$.

However, the quantum aspects of the Sp(2) theory were less addressed. The
operators which implement the BRST-anti-BRST symmetry at quantum level were
determined only in \cite{10}, using the ''collective fields'' as a basic
tool, but a more direct computation is not available. The well-defined
correlation functions and observables , the regularization and
renormalization programs were not developed in this context.

The aim of this paper is to use the technique mentioned above in order to
derive the expression of the quantum (anti) BRST operators in extended BRST
formalism (section 2), to explore the consequences concerning the
correlation functions (section 3) and to give an interpretation to the
consistency conditions obtained for the Sp(2) anomalies (section 4) showing
the equivalence with certain previous results.

\section{Deriving quantum (anti) BRST operators}

The main consequence of the relations (\ref{1})-(\ref{3}) is that the
path-integral: 
\begin{equation}
\label{4}Z=\int \left[ D\phi \right] U_K\exp \left( \frac i\hbar W\right) 
\end{equation}
does not depend on the choice of the gauge-fixing $K$, if the action $W$
satisfies the quantum master equation which may be written as: 
\begin{equation}
\label{5}\overline{\Delta }\left( \exp \left( \frac i\hbar W\right) \right)
=0
\end{equation}
and has been shown to be equivalent with the following two equations in
Sp(2) formalism: 
\begin{equation}
\label{6}\overline{\Delta }^a\exp \left( \frac i\hbar W\right) =0
\end{equation}
We use the well known denoting: 
\begin{equation}
\label{7}\overline{\Delta }=\overline{\Delta }^1+\overline{\Delta }^2
\end{equation}
\begin{equation}
\label{8}\overline{\Delta }^a=\Delta ^a+\frac i\hbar V^a
\end{equation}
The crucial requirement is now that a correlation functions: 
\begin{equation}
\label{9}I_K\left( B\right) \equiv \int \left[ D\phi \right] U_K\exp \left(
\frac i\hbar W\right) Y
\end{equation}
should not depend on the gauge-fixing $K$. This in turn is true only if $%
\exp \left( \frac i\hbar W\right) Y$ is ''divergencefree'', which means the
following condition: 
\begin{equation}
\label{10}\overline{\Delta }\left( \exp \left( \frac i\hbar W\right)
Y\right) =0
\end{equation}
must be fulfilled.

One can easily calculate the expression in (\ref{10}) and find: 
\begin{eqnarray}\label{11}
& & \Delta ^a\left( \exp \left( \frac i\hbar W\right) Y\right) = \nonumber \\
& & \left( -1\right) ^{\varepsilon _A}\frac{\delta _l}{\delta \phi ^A}
\frac \delta {\delta \phi _{Aa}^{*}}\left( e^
{\frac i\hbar W}Y\right) = \nonumber \\
& & e^{\frac i\hbar W}\left( -1\right) ^{\varepsilon _A}
\left( \left( \frac i\hbar \right) ^2\frac{\delta _lW}{\delta \phi ^A}
\frac{\delta W}{\delta \phi _{Aa}^{*}}Y+\frac i\hbar 
\frac{\delta _lW}{\delta \phi ^A}\frac{\delta Y}{\delta \phi _{Aa}^{*}}+\right.
\nonumber \\
& & \left. \frac i\hbar \frac{\delta ^2W}{\delta \phi ^A\delta \phi _{Aa}^{*}}Y+
\frac i\hbar \frac{\delta _lW}{\delta \phi _{Aa}^{*}}
\frac{\delta Y}{\delta \phi ^A}+\frac{\delta ^2Y}{\delta \phi ^
A\delta \phi _{Aa}^{*}}\right) = \nonumber \\
& & e^{\frac i\hbar W}\left[ \frac i\hbar \left[ \Delta ^aW+
\frac i{2\hbar }\left( W,W\right) ^a\right] Y+\right. \nonumber \\
& & \left. \left[ \Delta ^aY+\frac i\hbar \left( Y,W\right) ^a\right] \right] 
\end{eqnarray} which gives: 
\begin{eqnarray}\label{12}
& & \overline{\Delta }^a\left( e^{\frac i\hbar W}Y\right) =
\left( \Delta ^a+\frac i\hbar V^a\right) \left( e^{\frac i\hbar W}Y\right) =
\nonumber \\
& & e^{\frac i\hbar W}\left[ \frac i\hbar \left[ \overline{\Delta }^aW+
\frac i{2\hbar }\left( W,W\right) ^a\right] Y+\right. \nonumber \\
& & \left. \left[ \overline{\Delta }^aY+\frac i{\hbar }
\left( Y,W\right) ^a\right] \right] 
\end{eqnarray} If the quantum master equations are true ($a=1,2$), it
remains to require: 
\begin{equation}
\label{13}\overline{\sigma }^aY=0 
\end{equation}
and thus: 
\begin{equation}
\label{14}\overline{\sigma }Y=\left( \overline{\sigma }^1+\overline{\sigma }%
^2\right) Y=0 
\end{equation}
where: 
\begin{equation}
\label{15}\overline{\sigma }^aY\equiv \left( Y,W\right) ^a-i\hbar \overline{%
\Delta }^aY 
\end{equation}

We thus obtained the explicit expressions of the operators $\overline{\sigma 
}^a$ which obviously tend to the classical BRST anti BRST operators: 
\begin{equation}
\label{16}s^a=\left( ,S\right) ^a+V^a 
\end{equation}
when the limit $\hbar \rightarrow 0$ is considered.

\section{Correlation functions}

We may now investigate the significance of the condition (\ref{14}).

A first remark is that if we use a loopwise expansion for: 
\begin{equation}
\label{17}Y=\sum\limits_{p=0}^\infty \hbar ^pY_p
\end{equation}
and 
\begin{equation}
\label{17a}W=S+\sum\limits_{p=1}^\infty \hbar ^pM_p
\end{equation}
then the following set of relations is encoded in (\ref{14}): 
\begin{equation}
\label{18}\left( Y_0,S\right) ^a+V^aY_0\equiv s^aY_0=0
\end{equation}
\begin{equation}
\label{19}s^aY_1=i\Delta ^aY_0-\left( Y_0,M_1\right) 
\end{equation}
\begin{equation}
\label{20}s^aY_p=i\Delta ^aY_{p-1}-\sum\limits_{q=0}^{p-1}\left(
Y_q,M_{p-q}\right) ,\ { }p\geq 2
\end{equation}
One can see that the ''classical'' part of the quantum operator $Y$ is a
classical BRST-anti-BRST invariant.

A second remark concerns the possibility of finding ''trivial'' operators
which would generate zero correlation functions.

Let us assume that $\overline{\sigma }Y=0$ and: 
\begin{equation}
\label{21}Y=\overline{\sigma }X
\end{equation}
Then the corresponding correlation function is: 
\begin{eqnarray}\label{22}
& & I_K\left( Y\right) =\int \left[ D\phi \right] U_K
\exp \left( \frac i\hbar W\right) Y= \nonumber \\
& & \int \left[ D\phi \right] \left( \exp \left( \frac i\hbar W\right) 
\overline{\sigma }X\right) +\int \left[ D\phi \right] \overline{\Delta }J
\left( \exp \left( \frac i\hbar W\right) \overline{\sigma }X\right)  
\end{eqnarray} according to (\ref{3}). The second term in (\ref{22}) is zero
due to the property 
\begin{equation}
\label{23a}\int \left[ D\phi \right] \overline{\Delta }A=0
\end{equation}
(see \cite{8}). It remains: 
\begin{eqnarray}\label{23}
& & \int \left[ D\phi \right] \left( \exp \left( \frac i\hbar W\right) 
\overline{\sigma }X\right) =\int \left[ D\phi \right] \overline{\Delta }
\left( \exp \left( \frac i\hbar W\right) X\right) - \nonumber \\
& & \int \left[ D\phi \right] \exp \left( \frac i\hbar W\right) 
\left\{ \frac i\hbar \left[ \overline{\Delta }W+\frac i{2\hbar }
\left( W,W\right) \right] Y\right\}  
\end{eqnarray} which is zero due the same (\ref{23a}) and quantum master
equation.

We are led to the conclusion that in order to obtain good $I_K\left(
Y\right) $ we must use operators which satisfy $\overline{\sigma }Y=0$ but
are not $\overline{\sigma }$ - exact (there is no $X$ such that $Y=\overline{%
\sigma }X$).

On the other hand, the result: 
\begin{equation}
\label{24}\int \left[ D\phi \right] U_K\exp \left( \frac i\hbar W\right) 
\overline{\sigma }X=0
\end{equation}
for various functionals $X$ produces identities among correlation functions.
They embody the Ward identities associated to the extended BRST invariance
of the theory and with a particular choice of $X$ one can obtain, up to
normalization factors, the associated Schwinger-Dyson equations.

\section{Consistency conditions for Sp(2) anomalies and discussions}

The violation of the quantum master equation can be expressed as: 
\begin{equation}
\label{25}{\cal A}^a=\overline{\Delta }^aW+\frac i{2\hbar }\left( W,W\right)
^a
\end{equation}
which is then found to respect the ''consistency condition'' encoded in: 
\begin{equation}
\label{26}\overline{\sigma }^{\{a}{\cal A}^{b\}}=0
\end{equation}
and to be equivalent with the following set of equations: 
\begin{equation}
\label{26a}\left( {\cal A}_1^{\{a},S\right) ^{b\}}=iV^{\{a}\left(
M_1,S\right) ^{b\}}
\end{equation}
\begin{eqnarray}\label{28}
& & \left( {\cal A}_p^{\{a},S\right) ^{b\}}=iV^{\{a}
\left( M_p,S\right) ^{b\}}-\sum\limits_{n=1}^{p-1}
\left( {\cal A}_{p-n}^{\{a},M_n\right) ^{b\}}+
i\Delta ^{\{a}{\cal A}_{p-1}^{b\}}+ \nonumber \\
& & iV^{\{a}\sum\limits_{n=1}^{p-1}\left( M_{p-n},M_n\right) ^{b\}}\;,\;p\geq 2 
\end{eqnarray} if we expanded:

\begin{equation}
\label{29}{\cal A}^a=\sum\limits_{p=0}^\infty \hbar ^p{\cal A}_p^a
\end{equation}
and the definition (\ref{25}) was explicitly developed as: 
\begin{equation}
\label{30}{\cal A}_0^a=\frac 12\left( S,S\right) ^a+V^aS=0
\end{equation}
\begin{equation}
\label{31}{\cal A}_1^a=\Delta ^aS+i\left( M_1,S\right) ^a+iV^aM_1
\end{equation}
\begin{equation}
\label{32}{\cal A}_p^a=\Delta ^aM_{p-1}+i\left( M_p,S\right)
^a+iV^aM_p+i\sum\limits_{q=1}^{p-1}\left( M_q,M_{p-q}\right) ^a
\end{equation}
The property (\ref{26}) can be straightforwardly proven by using the
definition (\ref{25}) and the operator algebra for $\Delta ^a,V^a$ given in 
\cite{4}-\cite{6}.

A direct consequence is that:

\begin{equation}
\label{33}s^{\{a}{\cal A}_1^{b\}}=0
\end{equation}
\begin{equation}
\label{34}s^{\{a}{\cal A}_p^{b\}}=-\sum\limits_{n=1}^{p-1}\left( {\cal A}%
_{p-n}^{\{a},M_n\right) ^{b\}}+i\Delta ^{\{a}{\cal A}_{p-1}^{b\}},p\geq 2
\end{equation}
and one may easily see that (\ref{33}), the one loop order condition, is
exactly the one which has been previously deduced in a completely different
way in \cite{11}-\cite{13}, where it was written as: $sa^{-1}+\overline{s}%
a^1=0$..

We analyzed the quantifization of the Sp(2) - symmetric Lagrangian
formalism, underlying the derivation of the correct quantum operators in
this context, the relation with the standard case and the implications for
the correlation functions and anomalous terms. We should also stress that
the expressions involved in (\ref{6}) or in (\ref{25}) are not well-defined
unless an appropriate regularization scheme is used and this is a topic to
be covered in the near future.

\end{document}